\documentclass[twocolumn,pra,aps,longbibliography]{revtex4-2}

\usepackage{amsmath,amsfonts,amssymb,graphicx,hyperref,epstopdf,xcolor,tikz,scalerel,natbib,array}
\usepackage{mathptmx,braket}
\usepackage[T1]{fontenc}
\usepackage[utf8]{inputenc}
\usepackage{multirow}
\usepackage{array}
\DeclareMathAlphabet{\mathpcal}{OMS}{zplm}{m}{n}
 
\definecolor{lime}{HTML}{A6CE39}
\DeclareRobustCommand{\orcidicon}
{
	\begin{tikzpicture} 
	\draw[lime, fill=lime] (0,0) circle [radius=0.15] node[white] {{\fontfamily{qag}\selectfont \tiny ID}};
	\draw[white, fill=white] (-0.0625,0.095) 	circle [radius=0.007];
	\end{tikzpicture}
	\hspace{-2.2mm}
}
\newcommand\orcidID[1]{\href{https://orcid.org/#1}{\orcidicon}}

\newcommand{\Pdd}[2]{\frac{\partial#1}{\partial #2} }
\newcommand{\be}{\begin {equation}}
\newcommand{\ee}{\end {equation}}
\newcommand{\beqa}{\begin {eqnarray}}
\newcommand{\eeqa}{\end {eqnarray}}
\newcommand{\mb}{\mathbf}

\newcommand{\Sch}{Schr\"odinger }
\newcommand{\Exp}[1]{\text{e}^{#1}}
\newcommand{\phink}{\ket{\phi_k^n}}
\newcommand{\Enk}[1]{E_k^#1}
\newcommand{\psik}{\ket{\psi_k(t)}}
\newcommand{\alnk}[1]{\alpha_k^#1(t)}
\newcommand{\VB}[1]{VB$_#1$}
\newcommand{\CB}[1]{CB$_#1$}
\hypersetup{colorlinks,citecolor=blue,filecolor=black,linkcolor=blue,urlcolor=blue}

\begin{document}
 
\title{Role of inter- and intraband current in laser interaction with\\ bi-chromatic quasi-periodic crystals} 

\author{Amol R. Holkundkar\orcidID{0000-0003-3889-0910}}
\email[E-mail: ]{amol@holkundkar.in}
\author{Nivash R}
\author{Jayendra N. Bandyopadhyay\orcidID{0000-0002-0825-9370}}
 
\affiliation{Department of Physics, Birla Institute of Technology and Science - Pilani, Rajasthan,
333031, India}
\date{\today}

\begin{abstract}

We study the role of the inter- and intraband current in the laser interaction with the bi-chromatic quasi-periodic crystals. The interaction dynamics are simulated by solving the time-dependent \Sch equation in the $k$-space, and time evolution of the inter- and intraband current is obtained in a gauge invariant form. We observed that for certain bi-chromatic potential ratios, the energy band structure of the `valence band' and the `conduction band' 
facilitate the interband transitions only at the center or at the edge of the Brillouin zone, which leads to a very interesting population transfer mechanism between the bands. The temporal profile of the inter- and intraband current gives a detailed account of the interaction. The higher-order harmonic generation (HHG) is also studied for these bi-chromatic optical lattices, and the resultant harmonic yield is commented upon.     

\end{abstract}

\maketitle
 
\section{Introduction}
 
High-harmonic generation (HHG) from solids is gaining consistent traction and a field of contemporary interest around the globe because of the applications it promises in strong-field and attosecond physics. Though the HHG by the atomic gases formed the basis for the attosecond science,   the necessity of the complex setups with vacuum environments and sophisticated optics, along with the lower conversion efficiency of the HHG, poses significant challenges from the applied view. Solid-state HHG, from this perspective, has simplified operational details, with lower laser intensities and strong electron dynamics within the bands, which is remedied after the advent of the HHG by the  Bloch oscillations in the solids \cite{Ghimire2011,ghimire2012generation, ghimire2019high,wu2015high, you2016anisotropic,Ndabashimiye2016}. As a result, solid-state HHG promises a  compact source of the XUV radiations and attosecond spectroscopy \cite{Luu2015,RevModPhys.90.021002,vampa2017merge,PhysRevA.97.011401,Yue_22_josab}. The pioneering work on the HHG in the bulk ZnO crystal \cite{Ghimire2011} has opened the avenues in this vast field, and later the HHG is demonstrated in a wide range of materials such as larg-bandgap dielectrics \cite{Luu2015}, metasurface \cite{Liu2018}, graphene \cite{Yoshikawa2017_Science}, transition metal dichalcogenide \cite{Yoshikawa2019}, topological insulators \cite{Bai2021} and many more. For harmonic cutoff enhancement, yield, and optimization, numerous studies have reported wherein synthesized laser fields are used \cite{PhysRevA.101.033410,PhysRevA.102.063123,PhysRevA.96.043425}.

The solid-state HHG is mainly understood in terms of the delicate interplay between the two prominent physical mechanisms: interband polarization and the laser-driven intraband currents \cite{vampa2013theoretical,vampa2015linking, bian2018intra, Yoshikawa2019,PhysRevB.96.035112,Tancogne_Dejean2017, PhysRevA.98.023427, PhysRevB.96.035112, PhysRevB.98.235202,PhysRevA.97.013403}. It has been observed that the harmonics caused by the interband current always dominate the harmonics generated by the intraband current in the nonperturbative regime (harmonics above the minimum band gap energy), however in the perturbative regime, the intraband and interband harmonics are comparable with intraband harmonics being a slightly stronger \cite{vampa2013theoretical, PhysRevB.91.064302}. The gauge-independent (velocity or length) formulation of the inter- and intraband current for the HHG in the solids is presented in \cite{PhysRevB.96.035112,PhysRevB.98.235202}. The inter- and intraband transitions can also be understood from the motion of the Bloch electrons moving with the phase and group velocities in the coordinate space under the influence of the laser fields \cite{PhysRevA.97.013403}. 
 
The interband transitions typically happen when the electron passes through the region where the corresponding band gap between valence and conduction (or any other neighboring bands) is minimal, resulting in rapid changes in the electron population. In the context of the light-matter interaction, time-dependent population inversion between two energy levels is referred to as Rabi oscillations. The time evolution of the conduction band population in the context of the Rabi flopping/oscillations are previously reported \cite{PhysRevLett.116.197401,PhysRevB.82.075204}. The carrier-envelope phase (CEP) of the driving laser pulses are also found to significantly affect the band population \cite{PhysRevLett.116.197401, PhysRevLett.87.057401}. The control over the energy states of the electron using some external agency is desirable for understanding the underlying quantum dynamics. 

\begin{figure}[!t]
\begin{center}
 \includegraphics[totalheight=0.8\columnwidth]{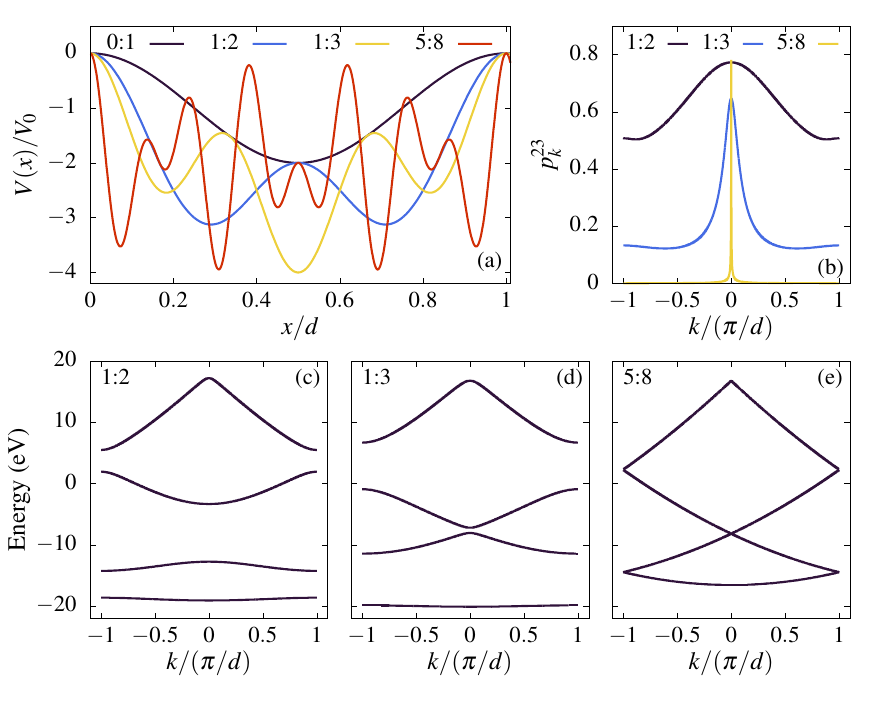}
  \caption{The normalized potential denoted by Eq. \eqref{potential} for 0:1, 1:2, 1:3 and 5:8 are presented in (a). The corresponding band structure showing the first four bands is shown for the ratios 1:2 (c), 1:3 (d), and 5:8 (e). The minimum bandgap between the \VB{2} and \CB{1} is 9.4 eV for 1:2 (c), 0.85 eV for 1:3 (d), and 0.012 eV for 5:8 ratio (e). The ratio 0:1 is pretty standard Mathieu-type potential \cite{PhysRev.87.807}, and hence the associated band structure is not presented. The $k$ dependent matrix element of transition from second band (\VB{2}) to the third band (\CB{1}) is presented in (b), i.e. $p_k^{23} = \braket{\phi_k^2|\hat{p}|\phi_k^3}$. }
\label{fig1}
\end{center}
\end{figure}

 For a given periodic crystal, the control over the dynamics or outcome of the solid-state HHG relies on tweaking the inter- and intraband currents using the synthesized laser fields. The band structure of the periodic crystal significantly affects the inter- and intraband current dynamics, and so the HHG \cite{PhysRevLett.118.087403}. The intraband current can have the clear signature of the band structure of the periodic crystal under study \cite{Lanin2017_Optica}. To this end, the optical lattices provide flexibility in terms of tweaking the lattice spacing and so the band structure \cite{PhysRevA.92.063426,Qiu2020}.
 
In this work, we have explored the interaction of the laser pulse with  quasi-periodic crystals and the role the inter- and intraband current plays in populating higher energy bands and  the HHG. We have solved the one-dimensional (1D) time-dependent \Sch equation (TDSE) in the quasi-momemtum space or $k$-space. The inter- and intraband current are calculated in a gauge invariant form along with the time-dependent population of the valence and conduction band. It is observed that the interference of the inter- and intraband current plays a very crucial role in suppressing high-energy harmonics. The paper is organized as follows: in Sec. \ref{sec2}, we discuss the theoretical and simulation details, followed by the results in Sec. \ref{sec3} and summary in Sec. \ref{sec4}. Throughout the manuscript, we have used the atomic units (a.u.) i.e., $|e|=m_e = \hbar = 1$.

\section{Theoretical methods}
\label{sec2}

\subsection{Bi-chromatic potential}
 
We study the interaction of the linearly polarized laser with the one-dimensional bi-chromatic periodic crystal. The laser polarization direction is considered to be along the optical lattice, and the bi-chromatic potential is modeled as \cite{PhysRevA.100.043420}: 
\be V(x) = -V_0 [A + B - A \cos(g\sigma_1 x) - B \cos(g\sigma_2 x)] \label{potential}\ee
here, $V_0$ denotes the depth of the potential, $g \equiv 2\pi/d$, $\sigma_1$ and $\sigma_2$ determines the respective frequencies of the bi-chromatic potential. Moreover, $d = 8$ a.u. is the lattice constant, $A$ and $B$ control the depth of the bi-chromatic potential. However, in our work we have set $A = B = 1$. There are two characteristic lengths in a unit cell: inter-atomic separation and the lattice constant $d$. In Fig. \ref{fig1}(a), we present the normalized potential for a unit cell with different frequency ratios $\sigma_1:\sigma_2$. The energy band structure for the frequency ratios 1:2, 1:3, and 5:8 is presented in Fig. \ref{fig1}(c), (d), and (e), respectively, where $V_0 = 0.3$ a.u. is used. Throughout the manuscript, we have considered that the electron is initially in the second band referred to as `Valence-Band' (\VB{2}). However, the higher lying bands will be referred to as `Conduction-Band' \CB{1} (third band), \CB{2} (fourth band) etc. The lower bandgap between \VB{2} and \CB{1} near $k =0$ for the ratio 5:8 makes it very interesting from the inter- and intraband current dynamics perspective, which we will discuss later in the paper. As can be understood from Fig. \ref{fig1}(a), the ratio 0:1 signifies 1 atom per cell, 1:2 denotes 2 atoms per cell, 1:3 denotes 3 atoms per cell, and 5:8 denotes 8 atoms per cell. The results are checked for convergence.  

\subsection{TDSE solver}
The electron wave function can be expanded in Bloch state basis $\phink$ for a particular value of the crystal quasimomentum $k$, and band index $n$ in order to numerically solve the TDSE in the velocity gauge \cite{korbman2013quantum,PhysRevA.108.063503}. The Bloch states are evaluated by solving the single-electron stationary \Sch equation with field-free Hamiltonian $\hat{H}_\text{o} = \hat{p}^2/2 + V(x)$:
\be \hat{H}_\text{o} \phink = \Enk{n} \phink.\ee
In position basis, the Bloch states can be written as:
\be \braket{x|\phi_k^n} \equiv \phi_k^n(x) = \sum_{\ell=1}^{N_\text{max}} C_{k,\ell}^n\ \Exp{i (k + 2\pi \ell/d) x},\ee
where $N_\text{max} = 19$ are used throughout the work and respective convergence is checked. The TDSE can be solved for electronic wavefunction $\psik$ as :
\be i \Pdd{}{t}\psik  = [\hat{H}_\text{o} + \hat{H}_\text{int}] \psik, \label{tdse_sol}\ee
where, $\hat{H}_\text{int} = \mb{A}(t) \cdot \hat{p}$ and $\mb{A}(t) \equiv -\int\limits^t \mb{E}(t') dt'$ is the vector potential associated with the laser pulse under dipole approximation with the electric field $\mb{E}(t) $ polarized along the $x$ direction as:
$ \mb{E}(t) = E_0 \sin^4(\pi t/T)\sin(\omega_0 t+\theta)\ \mb{e}_x,$
here, $E_0 \text{[a.u.]} \sim 5.342 \times 10^{-9} \sqrt{I_0}$ is the field amplitude with $I_0$ being the peak intensity of the pulse in W/cm$^2$, $T$ is the pulse duration, $\omega_0$ is the fundamental frequency of the laser pulse and $\theta$ is the carrier-envelope phase (CEP) of the pulse (considered to be zero unless stated otherwise).  
 Furthermore, at any given time instant, the time-evolving state $\psik$ can be expanded in the Bloch basis  
\be \psik = \sum_{n=1}^{N_\text{max}} \alnk{n} \phink, \label{psi_ex}\ee 
where $\alnk{n}$ are the time-dependent expansion coefficients. Using Eq. \eqref{psi_ex} in Eq. \eqref{tdse_sol}, the following coupled differential equations need to be solved \cite{korbman2013quantum}:
\be i \Pdd{\alnk{s}}{t}  = \Enk{s} \alnk{s} + A(t) \sum_{u=1}^{N_\text{max}} p_k^{su} \alnk{u} \label{coeff_sol}.\ee  
Here, $p_k^{su}$ is the matrix element of the momentum operator, which can be calculated as:
\be p_k^{su} = \braket{\phi_k^s|\hat{p}|\phi_k^u} = \sum_{\ell=1}^{N_\text{max}} (k + 2\pi\ell/d) \left(C_{k,\ell}^s\right)^{*} C_{k,\ell}^u.\ee
If we consider the electron initially in the band $q$, then the initial condition for solving Eq. (\ref{coeff_sol}) is $\alpha_k^s(0) = \delta_{qs}$. Finally, the single electron current density for a particular channel $k$ can be calculated as:
\be j_{ks}(t) = - \text{Re}[\braket{\psi_{ks}|\hat{p} + A(t)| \psi_{ks}}]. \label{current}\ee
In Eq. (\ref{current}), the subscript `$s$' denotes that the electron was in the band $s$ before the interaction.
Total current density can be calculated by summing over all the bands and integrating over the first Brillouin Zone (BZ) as:
\be j_\text{total}(t) = \sum_{s\in VB} \int j_{ks}(t) dk\label{current2}.\ee

\subsection{Band population calculation}

In order to estimate the Instantaneous Band Population (IBP) of the band $m$ at a given instant of time, the projection operator is expressed in terms of the field free Bloch basis \cite{PhysRevB.98.235202},
\be \hat{\Pi}_{mk} = \ket{\phi_k^m}\bra{\phi_k^m} \label{pimk}.\ee
In the velocity gauge the projection operator of Eq. \eqref{pimk} would transform as \cite{PhysRevA.81.063430}:
\be \hat{\Pi}_{mk}^\text{vel} = \text{e}^{-i A(t) \hat{x}}\ \hat{\Pi}_{mk}\ \text{e}^{i A(t) \hat{x}},\label{pimkt}\ee
where $A(t)$ is the vector potential associated with the driver field. The time-dependent population of the band $m$ is hence obtained by calculating the expectation value of the operator given in Eq. \eqref{pimkt} as \cite{PhysRevB.98.235202}:
\be \mathpcal{P}_m(t) = \int\limits_{BZ} \bra{\psi_k(t)} \hat{\Pi}_{mk}^\text{vel}  \ket{\psi_k(t)} dk.\label{bandpopu}\ee 


\subsection{Inter- and intraband current and HHG calculation} 

In Eq. \eqref{current2}, we obtained temporal dependence of the total current. To understand the inter- and intraband contribution in total current [Eq. \eqref{current2}], typically, one needs to rely on the Houston state basis \cite{wu2015high} when working in the velocity gauge; however, it has some numerical limitations to be used with a large number of bands. To remedy this, we used the Bloch state basis-based gauge-independent formulation, and accordingly, the inter- and intraband currents are calculated \cite{PhysRevB.98.235202} in terms of the population operator [Eq. \eqref{pimkt}]:
\be j_\text{inter}(t) = - \sum\limits_{m,m'\neq m} \iint_{BZ} dk dk'\ \hat{\Pi}_{mk}^\text{vel}\ (\hat{p} + A(t))\ \hat{\Pi}_{m'k'}^\text{vel},\ee  
\be j_\text{intra}(t) = - \sum\limits_{m} \iint_{BZ} dk dk'\ \hat{\Pi}_{mk}^\text{vel}\ (\hat{p} + A(t))\ \hat{\Pi}_{mk'}^\text{vel},\ee 
such that, $j_\text{total}(t) = j_\text{inter}(t) + j_\text{intra}(t)$.

The spectra of the emitted harmonics can be estimated by doing the Fourier transform of the current density and are given as: 
\be S_\text{total}(\omega) = \Big|\mathpcal{F}_\omega[j_\text{total}] \Big|^2 \ee 
where, $\mathpcal{F}_\omega[g(t)] = \int g(t) \exp[-i\omega t] dt$ is the Fourier transform of the time dependent function $g(t)$. However, it can also be interpreted as the summation of the harmonic spectra from the $j_\text{inter}(t)$, $j_\text{intra}(t)$ and the interference of both \cite{PhysRevA.98.023427} as follows:
\be S_\text{total}(\omega) = S_\text{inter}(\omega) + S_\text{intra}(\omega) + S_\text{intfer}(\omega) \ee 
where, 
\be S_\text{inter,intra}(\omega) = \Big|\mathpcal{F}_\omega[j_\text{inter,intra}]\Big|^2,\label{specinter}\ee
and,
\be S_\text{intfer}(\omega) = \mathpcal{F}_\omega^\ast[j_\text{inter}] \mathpcal{F}_\omega[j_\text{intra}] +  \mathpcal{F}_\omega^\ast[j_\text{intra}] \mathpcal{F}_\omega[j_\text{inter}],\label{intfer}\ee
denote the harmonic contribution from the interband, intraband current [Eq. \eqref{specinter}] and the interference of both is represented as Eq. \eqref{intfer}. The harmonic yield $Y$ for the frequency range $\omega_1$ to $\omega_2$ is calculated as: $Y = T^{-1} \int_{\omega_1}^{\omega_2}  S_\text{total}(\omega) d\omega$. Furthermore, the phase of the emitted harmonics is obtained as: $\varphi_i(\omega) = \rm{arctan2}[\rm{Im}\{S_i(\omega)\}, \rm{Re}\{ S_i(\omega)\}]$, with $i$ associated with either `inter', `intra' or `total' spectra.

\begin{figure}[!b]
\begin{center}
 \includegraphics[totalheight=0.8\columnwidth]{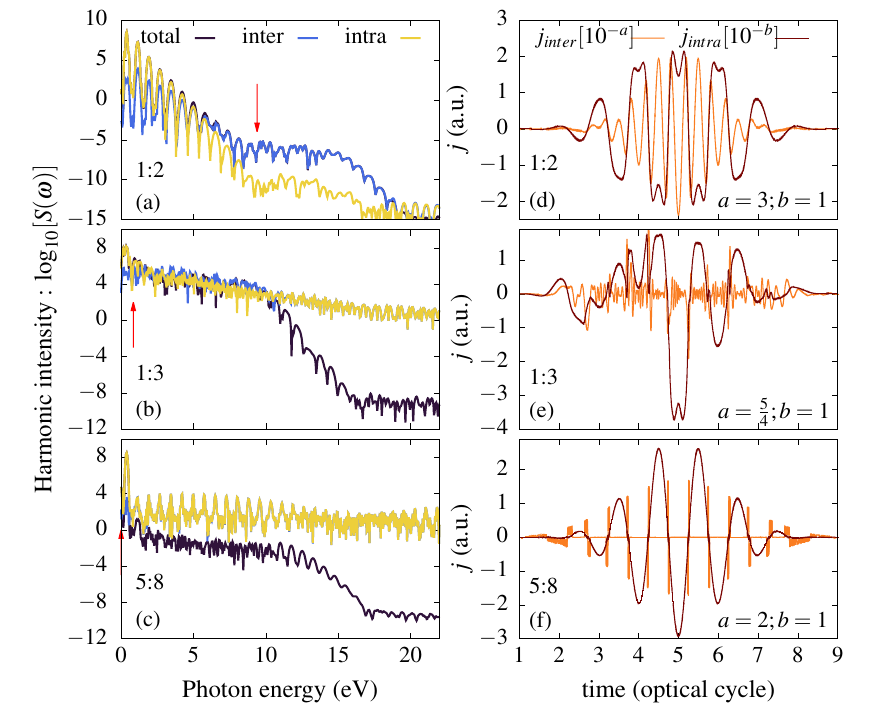}
  \caption{The HHG spectra for the frequency ratios 1:2 (a), 1:3 (b) and 5:8 (c) are presented for $S_\text{inter}$, $S_\text{intra}$, and $S_\text{total}$. The red arrow represents the minimum bandgap between the \VB{2} and \CB{1} for respective cases [refer Fig. \ref{fig1}]. The respective inter- and intraband currents are presented for 1:2 (d), 1:3 (e), and 5:8 (f). As mentioned, the inter- and intraband current in (d-f) is plotted on the same axis with different scaling parameters $a$ and $b$.}
\label{fig2}
\end{center}
\end{figure}

\section{Results and discussion}
\label{sec3}
 
\begin{figure}[!b]
\begin{center}
 \includegraphics[totalheight=0.8\columnwidth]{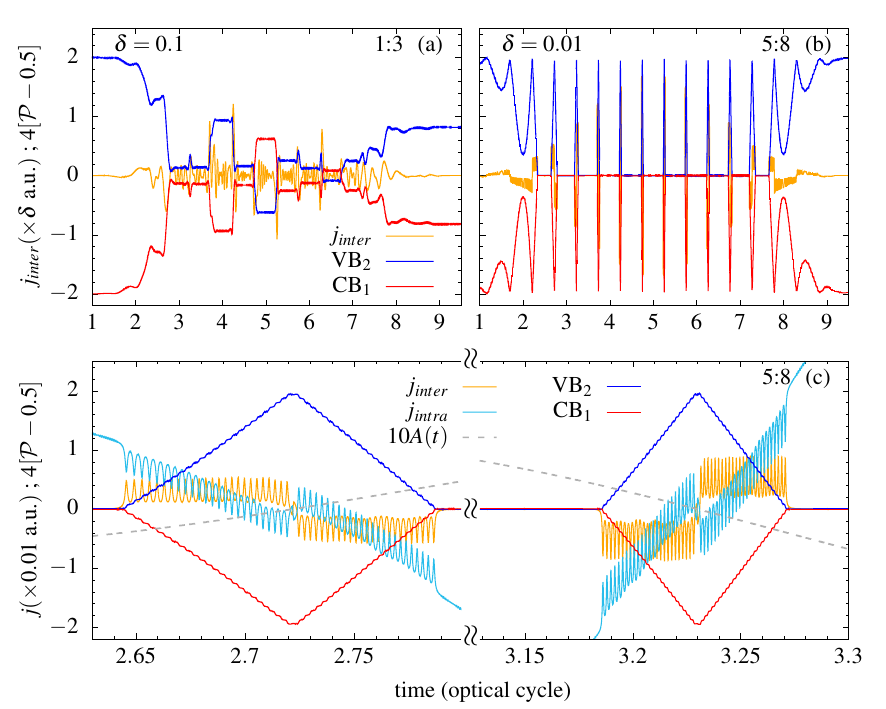}
  \caption{Temporal evolution of the interband current along with the \VB{2} and \CB{1} band population is presented for frequency ratio 1:3 (a) and 5:8 (b). The zoomed version of (b) in the range $2.6\tau \leq t \leq 3.3\tau$ is shown in (c), with $\tau$ being an optical cycle. In (c), the temporal profile of the vector potential (dashed line) is also shown along with the intraband current. In order to represent band population ($\mathpcal{P}$) and the interband current on the same scale, an appropriate scale factor $\delta$ is mentioned for interband current, and the quantity  $4 \mathpcal{P} - 2$ is plotted, such that $\mathpcal{P} = 0$ corresponds to $-2$ and $\mathpcal{P} = 1$ corresponds to $+2$ on the $y$-scale.}
\label{fig3}
\end{center}
\end{figure}
 
The interaction of the 10 cycles, 3.2 $\mu$m laser with a  peak intensity of $6\times 10^{11}$ W/cm$^2$ is considered with the quasi-periodic crystals with frequency ratios 1:2, 1:3, and 5:8, having $V_0 = 0.3$ a.u, and $10\%$ contribution around $k = 0$ in the BZ is considered i.e. $-0.1 < k d/\pi < 0.1$ (unless otherwise stated) in order to understand the underlying interaction dynamics.  

\subsection{HHG for different ratio $\sigma_1 : \sigma_2$}
The decomposition of the HHG spectra with inter- and intraband contributions is presented in Fig. \ref{fig2}(a-c). 
In the perturbative regime of the solid HHG, inter- and intraband HHG are dominant with intraband current, and so the harmonics are stronger. However, in the nonperturbative, regime mainly interband current contributes, though both the inter- and intraband currents show the plateau structures \cite{vampa2013theoretical,PhysRevA.98.023427,PhysRevA.97.013403}. Here, we can corroborate these findings in Fig. \ref{fig2}(a) for frequency ratio 1:2, wherein the below bandgap harmonics ($\lesssim 9.4$ eV) the intraband contribution ($S_\text{intra}$) is comparable to the interband ($S_\text{inter}$) one. However, in the range $\gtrsim 9.4$ eV,   $S_\text{inter}$ overlap   $S_\text{total}$ (total contribution), implying the interference ($S_\text{intfer}$) and the $S_\text{intra}$ terms are less significant.  Furthermore, the bichromatic frequency ratio is 1:3 [Fig. \ref{fig2}(b)] show a very crucial role played by the $S_\text{intfer}$ terms in the total spectra. It can be seen from Fig. \ref{fig2}(b) that in the perturbative regime ($\lesssim 0.85$ eV), the $S_\text{intra}$ is slightly dominant and in the nonperturbative regime up to harmonic cutoff ($\gtrsim 0.85$ eV and $\lesssim 10$ eV) the $S_\text{inter}$ is slightly stronger than the $S_\text{intra}$. However, in this case, $S_\text{intfer}$ is very strong and significantly affects the HHG process, as can be seen from Fig. \ref{fig2}(b) after the cutoff,   $S_\text{intfer}$  negates the contribution from $S_\text{inter}$ and  $S_\text{intra}$, giving a clear harmonic cutoff around $\sim 10$ eV. Moreover, the dominance of the interference term $S_\text{intfer}$ is found to be very stark for the case when the frequency ratio is changed to 5:8, i.e., Fig. \ref{fig2}(c). In this case, mostly all the HHG spectra is in the nonperturbative regime as the minimum bandgap of \VB{2} and \CB{1} is $\sim 0.01$ eV [refer Fig. \ref{fig1}]. It can be seen from Fig. \ref{fig2}(c) that the contribution of $S_\text{inter}$ and $S_\text{intra}$ are exactly identical and the contribution from   $S_\text{intfer}$ term reduces the combined contribution of $S_\text{inter} + S_\text{intra}$ significantly and give rise to the clean harmonic cutoff at $\sim 12$ eV. Later, we will discuss how the minute phase difference between $S_\text{inter}$ and $S_\text{intra}$ exactly mimics the total HHG obtained and reinforces the importance of the interference of the inter- and intraband currents.     
 
The inter- and intraband current for all these three cases is also illustrated in Fig. \ref{fig2}(d-f) and for visual appeal $j_\text{inter}$ and $j_\text{intra}$ are plotted on the same scale with appropriate scaling factors. It should be worth noting that as the minimum bandgap reduces for cases from Fig. \ref{fig2}(a) - (c), the interband currents gain very fast Rabi-like oscillations, which effectively cause the rapid oscillations of the population between \VB{2} and \CB{1}. As a result, the distinction between   $S_\text{inter}$ and $S_\text{intra}$ (which are Fourier transforms of the $j_\text{inter}(t)$ and $j_\text{intra}(t)$) is completely lost in Fig. \ref{fig2}(c), because say the fast depletion in one band is fast accumulation in another band in time domain. Hence, in the spectral domain, these two events are the same. These fast oscillations in the interband current [Fig. \ref{fig2}(f)] occurs near the point where the vector potential $A(t) \sim 0$, which happens near $k \sim 0$. In order to further elucidate this fast oscillating nature of the inter- and intraband current near zero crossing of the vector potential, we have presented the zoomed version of Fig. \ref{fig2}(f) along with the respective band population in Fig. \ref{fig3}.   
 
The detailed inter- and intraband current features for the frequency ratio 1:3 and 5:8 are presented in Fig. \ref{fig3}. The band population is calculated using Eq. \eqref{bandpopu}, and it can be seen from the temporal profile of the band population that, the dynamics are effectively reduced to two band problem for the given laser and potential parameters. If we compare the interband current of 1:3 and 5:8 frequency ratio in Fig. \ref{fig3}(a) and (b), then it is observed that the band population perfectly oscillates between \VB{2} and \CB{1} for 5:8 frequency ratio, the feature which is missing for 1:3 ratio (later we will discuss this aspect). The zoomed version of Fig. \ref{fig3}(b) is illustrated as Fig. \ref{fig3}(c) and following observations can be made from the same:
\begin{enumerate}
  \item During the complete cycle of the pulse, these rapid oscillations in the interband current happen near the point where $A(t) = 0$, which corresponds to the electric field maxima for that particular cycle in the pulse. 
  \item There are two parts in each cycle; in the first half, the population is transferred from \CB{1} to \VB{2}, and in the second half, the reverse happens, and population transfers from \VB{2} to \CB{1}. In between successive cycle, the \VB{2} and \CB{1} are equally populated with $\mathpcal{P} = 0.5$.   
  \item The rapid time variation of the intraband current is out of phase with the interband current.
  \item These time variations of the interband currents only last till the population transfers complete from \VB{2} to \CB{1} and vice-versa. 
  \item The time interval for a complete cycle of populating and then depopulating the \CB{2} reduces in the window $3.15\tau < t < 3.3\tau$ (with $\tau$ being one optical cycle) as compared to the time window $2.6\tau < t < 2.8\tau$.
  \item All the above points can easily be understood in terms of the Rabi oscillations. The Rabi frequency is given by:
  \be \Omega_R(k,t) \propto |\mb{E}(t)| \frac{\braket{\phi_k^\text{\CB{1}}|\hat{p}|\phi_k^\text{\VB{2}}}}{E_k^\text{\CB{1}} - E_k^\text{\VB{2}}} \label{rabifreq}\ee   
  The strong field amplitude in the window $3.15\tau < t < 3.3\tau$ manifests in higher Rabi frequency, and hence the population transfer from the \VB{2} to \CB{1} and vice-versa takes less time. These features are consistent if we see the time dependence over a complete laser pulse duration [Fig. \ref{fig3}(b)].    
  \item The transition matrix element between \VB{2} and \CB{1}, along with the bandgap between the two for particular $k$ values, also plays a very crucial role in determining the Rabi frequency. In this particular case of 5:8 ratio, we see that the transition matrix element in the momentum space $\braket{\phi_k^\text{\CB{1}}|\hat{p}|\phi_k^\text{\VB{2}}}$ only peaks near minimum bandgap only, i.e. at $k = 0$ [refer Fig. \ref{fig1}(b)]. However, for 1:3 ratio the contribution from the neighboring Bloch states also plays a role and so the interband current and the band population is not similar to as it is with 5:8 ratio. 
\end{enumerate} 

\begin{figure}[t]
\begin{center}
 \includegraphics[totalheight=0.8\columnwidth]{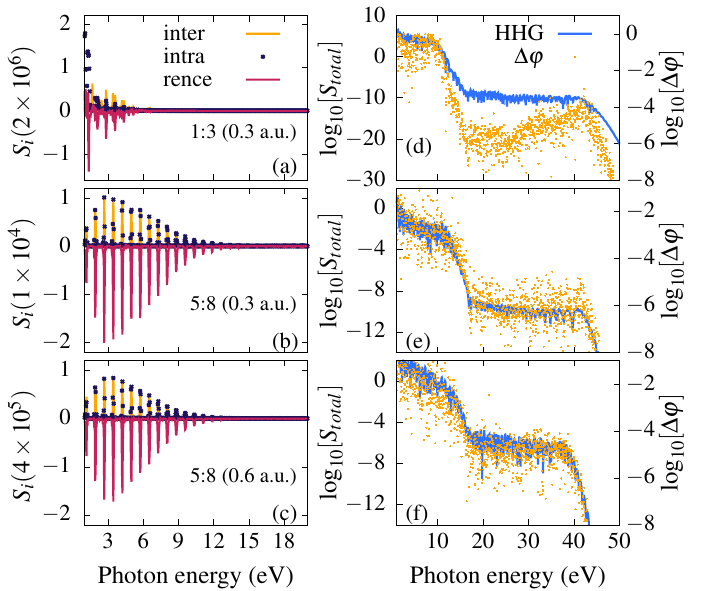}
  \caption{Harmonic spectra associated with the interband, intraband, and interference terms [refer Eq. \eqref{specinter} and \eqref{intfer}] are presented for the bichromatic ratio 1:3 ($V_0 = 0.3$ a.u.) (a), 5:8 with $V_0 = 0.3$ a.u. (b) and $V_0 = 0.6$ a.u. (c). On the right panel, the respective total HHG spectra are plotted in (d)-(f). The phase difference $\Delta \varphi$ [refer Eq. \eqref{varphi}] is also shown in the figure.     }
\label{fig4}
\end{center}
\end{figure}

\subsection{Interference of inter- and intraband currents}
We learned from the Figs. \ref{fig2} and \ref{fig3}, how the electric field amplitude and the matrix element of the dipole operator affect the Rabi oscillations between the \VB{2} and \CB{1}, which eventually affect the respective band-population. In order to further understand the nature of the HHG spectra and the role of interference of inter- and intraband contribution, the HHG spectra for 1:3 and 5:8 cases are shown in Fig. \ref{fig4}. It can be understood from this figure that for the case of 5:8 ratio, the inter- and intraband harmonics `almost' overlap with each other, and the interference between the two causes the emergence of the total HHG spectra with multiple plateaus. As we saw previously in Fig. \ref{fig3}, the population transfer in case of 5:8  occurs only near the peak of the electric field (minimum of vector potential), and as a result, the process repeats in each half cycle, and hence  only the odd order harmonics are prominently visible for inter- and intraband HHG in Fig. \ref{fig4}(b) and (c) for 5:8 case. However, for the 1:3 case, no such observation is made [refer Fig. \ref{fig3}(a)]. The amplitude of the interference term is of the order of addition of inter- and intraband harmonics, hinting towards strong destructive interference, and as a result, the clear odd order harmonics are not visible in total HHG spectra in Fig. \ref{fig4}(d)-(f). Furthermore, the phase difference between the inter- and intraband harmonics is defined as:
\be \Delta \varphi = 1 - \frac{1}{\pi} \Big|\varphi_\text{inter} - \varphi_\text{intra}\Big|\label{varphi}.\ee 
The offset from perfect destructive interference ($\Delta \varphi = 0$) between the inter- and intraband harmonics is observed to accurately manifest in the observed cutoff energies of full HHG spectra.   
  
\begin{figure}[b]
\begin{center}
 \includegraphics[totalheight=0.5\columnwidth]{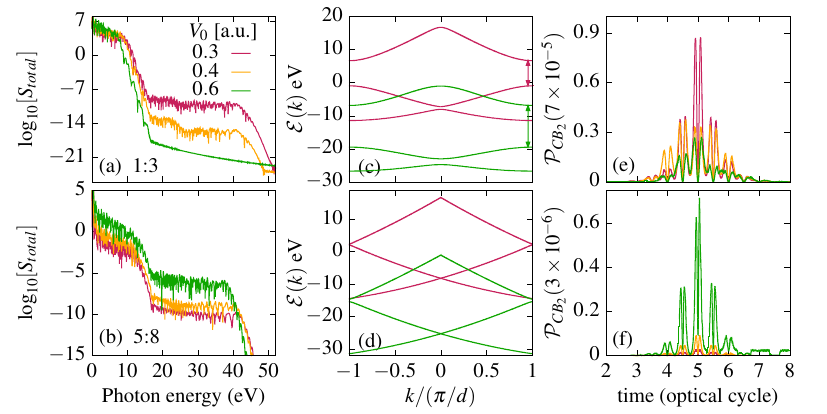}
  \caption{Harmonic spectra for 1:3 and 5:8 cases are illustrated in (a) and (b) respectively for different values of the $V_0$ [refer Eq. \eqref{potential}]. The energy bands \VB{2}, \CB{1} and \CB{2} for $V_0 = 0.3$ and 0.6 a.u. cases are illustrated for 1:3 (c) and 5:8 ratios (d). The temporal evolution of the population of \CB{2} i.e. $\mathpcal{P}_{CB_2}$, are compared for $V_0 = 0.3$ and 0.6 a.u. cases for 1:3 (e) and 5:8 (f) ratios.}
\label{fig5}
\end{center}
\end{figure}

\subsection{Effect of potential depth $V_0$ on the HHG spectra}
So far, we learned that the typical characteristics of the 5:8 ratio make it interesting from the interaction perspective, as the transition to higher bands happens either at the center of BZ or at the edge of the same. In order to further elucidate on this aspect, in Fig. \ref{fig5}, we have presented the HHG spectra for 1:3 and 5:8 cases for different potential depth $V_0$ [refer Eq. \eqref{potential}]. It can be seen from Fig. \ref{fig5}(a) and (b) that with the increase in the potential depth, the efficiency of the secondary plateau $\sim 20 - 40$ eV decreases for 1:3 ratio and correspondingly increases for 5:8 ratios. This can be understood by studying the energy bands for 1:3 and 5:8 ratios for different $V_0$ values. We have presented the \VB{2}, \CB{1} and \CB{2} energy bands for both 1:3 and 5:8 cases in Fig. \ref{fig5}(c) and (d) for $V_0 = 0.3$ and 0.6 a.u. The photons in the energy range $\sim 20 - 40$ eV (which comprises the secondary plateau) are mostly emitted by the transition from the \CB{2} to the lying band, and hence the time-dependent band-population of \CB{2} can be a good measure to co-relate to the harmonic efficiency of the radiation in the secondary plateau. It can be seen that for the case of 1:3 ratio, the energy difference between the \CB{1} and \CB{2} increases from $\sim 7.6$ eV to  $\sim 12.6$ eV with an increase in the potential depth from 0.3 a.u. to 0.6 a.u., and so the instantaneous band population of \CB{2} reduces implying the less probability to Zener tunneling to \CB{2} [refer Fig. \ref{fig5}(e)]. On the contrary, for a 5:8 ratio, as we increase the $V_0$, the bands \CB{1} and \CB{2} flatten a bit at the edge of the BZ. As a result, the Zener tunneling increases, giving enhanced population for \CB{2}. This enhanced population of \CB{2} manifests in enhanced efficiency in HHG yield for the energy range $\sim 20 - 40$ eV.

\begin{figure}[t]
\begin{center}
 \includegraphics[totalheight=0.8\columnwidth]{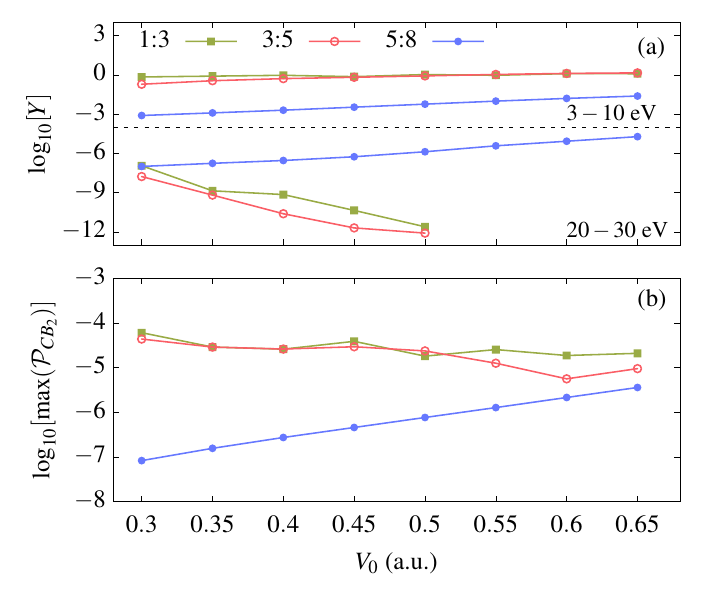}
  \caption{The variation of the harmonic yield with the potential depth $V_0$ is presented for three different bichromatic ratios in (a). However, the maximum of the band population $\mathpcal{P}_{CB_2}$ dependence on the $V_0$ is shown in (b). In (a), plots above (below), the black dashed line represents the harmonic yield in the energy range 3 - 10 eV (20 - 30 eV).  }
\label{fig6}
\end{center}
\end{figure}

In Fig. \ref{fig6}, the variation of the harmonic yield and maximum of the \CB{2} population with $V_0$ is presented for different bi-chromatic ratios. The harmonic yield in the energy range 20 - 30 eV is not plotted for 1:3, and 3:5 ratio for $V_0 > 0.5$ a.u. [Fig. \ref{fig6}(a)], as the secondary plateau completely diminishes for $V_0 > 0.5$ a.u. It can be understood from this figure that the harmonic yield in the energy range belonging to the secondary plateau increases (decreases) with $V_0$ which is reflected by the increase (decrease) in the population of the band, which contributes to the said energy range in HHG spectra [Fig. \ref{fig6}(b)]. The harmonics in the energy range 3-10 eV (primary plateau) arise mainly by the transition from the \CB{1} to \VB{2} in all the cases. For the 5:8 case, as the $V_0$ increases, the bandgap between \VB{2} and \CB{1} increases, and the band slopes near $k = 0$ reduce. This combined effect causes enhanced Zener tunning to higher bands, enhancing harmonic yield in the 3-10 eV range and the secondary plateau by subsequent promotion to \CB{2}. 

\begin{figure}[b]
\begin{center}
 \includegraphics[totalheight=0.8\columnwidth]{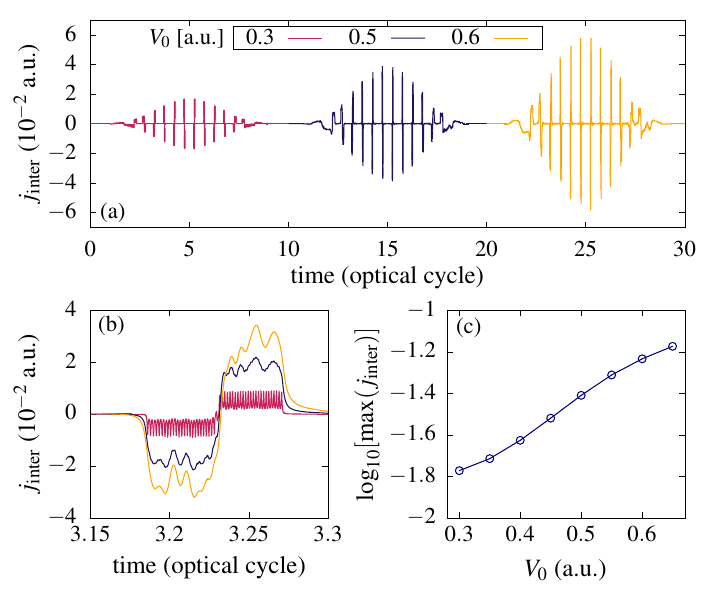}
  \caption{Temporal dependence of $j_\text{inter}$ is presented three different values of $V_0$ for 5:8 ratio case (a). The zoomed version of (a) is illustrated in (b), and the variation of a maximum of $j_\text{inter}$ with $V_0$ is shown in (c). It should be noted that for visual appeal, the temporal profile in (a) for $V_0 = 0.5$ a.u. and 0.6 a.u. are intentionally shifted by 10 and 20 cycles, respectively.}
\label{fig7}
\end{center}
\end{figure}

We learned that for the case of 5:8 ratio, the Zener tunneling to higher bands only happens either at the center or at the edge of BZ, which is also quite visible from the temporal evolution of the interband current $j_\text{inter}(t)$ as discussed in detailed in Fig. \ref{fig3}. This fact brings us to a question, how does the strength of   $j_\text{inter}$ affect the HHG yield? In order to explore this facet, in Fig. \ref{fig7} we have analyzed   $j_\text{inter}$ for different potential depths $V_0$. It can be observed from Fig. \ref{fig7}(a) and (b) that as we increase   $V_0$, the peak value of the interband current increases; this is due to the enhanced Zener tunneling among bands [lower `velocity' $\sim \nabla_k \mathpcal{E}(k)$ near the edge of the BZ] with increase in $V_0$ for 5:8 ratio. Furthermore, as   $V_0$ increases, the band slope decreases, and so the neighboring $k$ values also contribute to the transition or in other words, the $p_k^{23}(k)$ broadens [refer Fig. \ref{fig1}(b)]. This results in the disappearance of Rabi-type fast oscillations, as we have seen for $V_0 = 0.3$ a.u. case. Typically, we observe these Rabi type of oscillations when mostly only two energy levels are involved near the $k=0$. However, as more and more $k$ values are participating in the transition, this assembly of `two-level' system at $k = 0$ breaks down and rapid oscillations are replaced by the strong interband current. The strong interband current eventually populates higher conduction bands, enhancing the harmonic yield in the secondary plateau region. However, it should be noted that even in this scenario, the interband current only contributes near the peak of the electric field or the temporal points where $A(t) \sim 0$. The maximum interband current variation with the potential depth is also presented in Fig. \ref{fig7}(c); it is observed that the maximum interband current is positively correlated with the harmonic yield of the secondary plateau [Fig. \ref{fig6}(a)]. In order to further understand the HHG spectra and the role of interband current, next we discuss the time-frequency analysis for two cases of 5:8 with $V_0 = 0.3$ and 0.6 a.u.

\begin{figure}[t]
\begin{center}
 \includegraphics[totalheight=0.5\columnwidth]{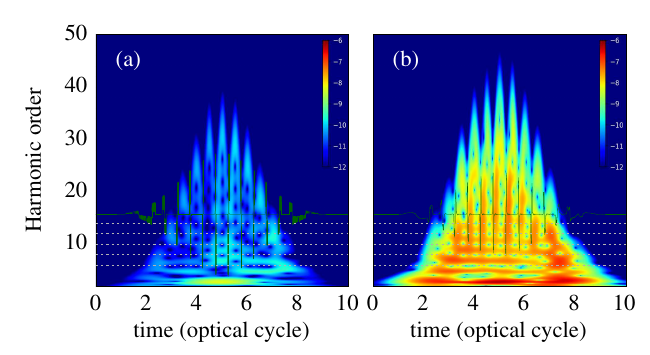}
  \caption{The time frequency analysis for 5:8 case with $V_0 = 0.3$ a.u. (a) and 0.6 a.u. (b) are shown. In both the figures, the horizontal dashed gray lines show even harmonics. The temporal profile of the $j_\text{inter}$ is also superimposed on these plots. }
\label{fig8}
\end{center}
\end{figure}


\subsection{Time frequency analysis of HHG}

As we discussed in Figs. \ref{fig3} and \ref{fig4} that the peculiar feature  the  bi-chromatic ratio 5:8 is that the transition to higher bands only happens at the center or at the edge of the BZ, which implies that if you consider the bands \VB{2} and \CB{1} (corresponds to the primary plateau $\lesssim 15$ eV) the interband transition will only happen near $k \sim 0$ as $p_k^{23}$ is maximum only for $k = 0$ [refer Fig. \ref{fig1}(b)]. This condition will be met only when $A(t) \sim 0$ and the transition is facilitated, giving rise to $j_\text{inter}$. This process will happen at each half cycle, giving rise to the only odd-order harmonics as seen in Fig. \ref{fig4}(b) and (c), however due to the interference between the inter- and intraband harmonics, clear odd-order harmonics are not that prominent [refer Fig. \ref{fig4}(e) and (f)]. In Fig. \ref{fig8} we have presented the time-frequency analysis of the HHG spectra for a 5:8 ratio for two different values of $V_0$. It is observed that even in the total HHG spectra, the intensity of the even harmonics is significantly reduced, as shown by the dashed horizontal lines in Fig. \ref{fig8}(a) and (b). We have also superimposed the temporal profile of the interband current and it can be seen that the intensity of the harmonics between two consecutive peaks of $j_\text{inter}$ is 
smaller than the one observed at peaks of $j_\text{inter}$. This further establishes the fact that the interband current plays a prominent role in the HHG process from the (quasi) periodic crystals and reduction of the same 
is reflected in weaker harmonic efficiency. These missing even harmonics are not observed for 0:1, 1:3, and 3:5 cases, as the interband current in those scenarios is the superposition of all different channels corresponding to different $k$ values. 
 
\section{Summary}
\label{sec4}

In summary, we have studied the role of inter- and intraband current on the HHG by the laser interaction with the bi-chromatic quasi-periodic crystals with the frequency ratios $\sigma_1:\sigma_2$ being 1:3, 3:5, and 5:8 [Eq. \eqref{potential}]. It is observed that the typical characteristics of the energy bands associated with the 5:8 ratio make it possible to have the transition from the \VB{2} to \CB{1} only at the center or at the edge of the BZ, which leads to very interesting population transfer mechanism between \VB{2} and \CB{1}. The effect of the potential depth on the generated harmonics and the interband current is also studied, and it is observed that the harmonic yield of the secondary and primary plateau increases with increasing the potential depth for a 5:8 ratio, which is positively correlated to the strength of the interband current. The conclusions are vindicated by also studying the time-dependent instantaneous band populations. The presented analysis is also observed to be true even when the decoherence effects are included in a phenomenological manner \cite{PhysRevA.108.063503, PhysRevA.103.063109}. The laser field parameters and their respective influence on the HHG by these quasi-periodic crystals are beyond the scope of the current manuscript and are reserved for future studies. The temporal control of the band-population for such optical lattices might interest the `\textit{quantum-computing}' fraternity. 
 
\section*{Acknowledgments} The authors acknowledge the Science and Engineering Research Board (SERB), India, for funding Project No. CRG/2020/001020.


%
\end{document}